\documentclass[a4paper,fleqn]{cas-sc}

\usepackage[authoryear,longnamesfirst]{natbib}
\usepackage{amssymb}
\usepackage{graphicx}%
\usepackage{multirow}%
\usepackage{amsmath,amssymb,amsfonts}%
\usepackage{amsthm}%
\usepackage{mathrsfs}%
\usepackage[title]{appendix}%
\usepackage{xcolor}%
\usepackage{textcomp}%
\usepackage{manyfoot}%
\usepackage{booktabs}%
\usepackage{algorithm}%
\usepackage{algorithmicx}%
\usepackage{algpseudocode}%
\usepackage{listings}%

\usepackage{hyperref}   
\hypersetup{
	colorlinks=true,
	linkcolor=blue,
	citecolor=blue,
	urlcolor=blue,
}

\usepackage{mathtools}

\usepackage{lipsum}
\usepackage{diagbox}
\usepackage{indentfirst}

\usepackage{array}
\usepackage{tabularx}
\usepackage{kotex}

\def\tsc#1{\csdef{#1}{\textsc{\lowercase{#1}}\xspace}}
\tsc{WGM}
\tsc{QE}
\tsc{EP}
\tsc{PMS}
\tsc{BEC}
\tsc{DE}

\begin{document}
\let\WriteBookmarks\relax
\def\floatpagepagefraction{1}
\def\textpagefraction{.001}
\shorttitle{Leveraging social media news}
\shortauthors{S.Jin et~al.}

\title [mode = title]{A Global Routing Method for Electronic Circuit Design Integrating Flow Matching and Constrained A*}

\author[1]{Sanghoon Jin}
\ead{jsanghooni@kaist.ac.kr}
\credit{Conceptualization of this study, Data curation, Writing - Original draft preparation}
\author[2]{Sumin Lee}
\cormark[1]
\ead{sumlee@jbnu.ac.kr}
\credit{Methodology, Data curation, Writing - Original draft preparation}
\author[1,3]{Namwoo Kang}
\cormark[1]   
\ead{nwkang@kaist.ac.kr}
\credit{Conceptualization of this study, Data curation, Writing - Original draft preparation }

\affiliation[1]{Cho Chun Shik Graduate School of Mobility, Korea Advanced Institute of Science and Technology (KAIST), 193 Munji-ro, Yuseong-gu, Daejeon 34051, Republic of Korea}

\affiliation[2]{Advanced Transportation Machinery Research Center & Department of Mechanical System Engineering, Jeonbuk National University (JBNU), 567 Beakje-daero, duckjin-gu Jeonbuk 54896, Republic of Korea}

\affiliation[3]{Narnia Labs, 193 Munji-ro, Yuseong-gu, Daejeon 34051, Republic of Korea}

\cortext[cor1]{Corresponding author}

\begin{abstract}
Global routing is a fundamental issue in the design of electronic circuits, as it directly relates to the efficiency, reliability, and compactness of the overall layout of integrated systems. As the demand for higher functionality and reliability within a limited chip area increases, the importance of routing as a design consideration has never been greater. Efficient global routing optimization techniques are therefore required to connect all pins with minimal wirelength (WL), while avoiding congestion and interference across multiple nets within the design space.
As the number of nets to be routed increases, global routing inherently suffers from a combinatorial explosion in the design space, making it increasingly difficult to obtain feasible solutions. In this paper, we focus on solving constrained global routing problems based on a heuristic A*-based framework.
Based on this concept, a hybrid approach is proposed by integrating constrained A* with flow matching (FM). The method utilizes FM to construct a global cost map that captures routing tendencies across the design space, which is then used to guide the constrained A* search process. This enables efficient exploration of the solution space while satisfying routing constraints.
Experimental results demonstrate that the proposed method achieves comparable WL and via usage to baseline methods while maintaining constraint satisfaction across a wide range of scenarios. Notably, the framework shows strong capability in solving complex routing instances that are difficult to handle using conventional approaches. Additionally, the method shows robust generalization performance across different problem sizes and net configurations.

\end{abstract}

\begin{graphicalabstract}
\includegraphics{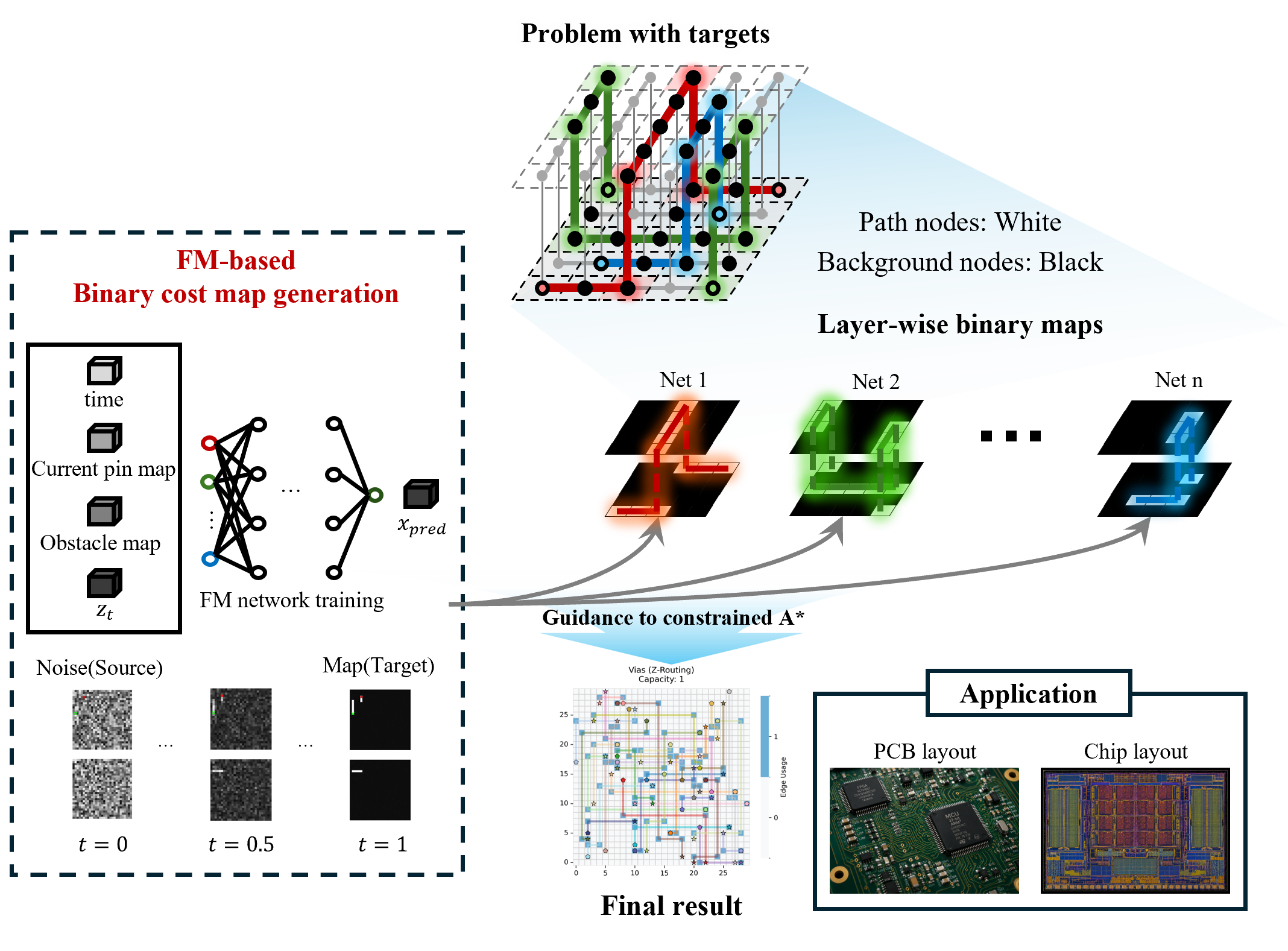}
\end{graphicalabstract}

\begin{highlights}
\item A novel hybrid framework that integrates flow matching (FM) with constrained A*-based global routing
\item The first work that introduces a binary flow matching formulation for cost map generation
\item demonstrating feasible and high-quality routing solutions in complex situations
\end{highlights}

\maketitle


\section{Introduction}\label{sec1}

As the demand of compactness in electronic devices grows, the complexity of integrated circuits (ICs) is increasing, and millions of devices are included in a single chip while the number is increasing. This means that the number of devices per unit area increases exponentially, and as a result, the need for routing problems that connect all of these devices is also increasing. IC routing is largely divided into global routing and detailed routing, of which global routing is a matter of dividing the entire chip space into grid space and then finding how to connect pins belonging to each cell (G-cell). This involves considering various conditions, such as the capacity of the edge connecting the G-cell, crossover between paths, and minimizing the use of vias across all paths. An important point to note is that even the problem of finding the shortest path that meets these constraints for only one 2-pin net is proven to be an NP-hard problem \citep{nphard}. The problem of global routing in real IC becomes an extremely difficult problem to perform for millions of pins and corresponding number of paths. These properties suggest the need to approach them from a different perspective than the well-known combinatorial optimization problems, such as traveling salesman problems (TSP).

Routing methodologies are generally categorized by their execution strategy into sequential and concurrent approaches. Sequential routing, exemplified by A*-based algorithms, connects nets one after another. This inherently leads to the net ordering problem, where prior connections restrict available resources for subsequent ones. While techniques like rip-up and reroute (RRR) are employed to mitigate this, they often suffer from convergence issues such as oscillation. Conversely, concurrent routing addresses multiple nets simultaneously for better resource allocation, but its computational complexity often renders it impractical for large-scale problems.

These structural challenges are further compounded by the limitations of conventional solving mechanisms. Traditional approaches—ranging from manual expert design to heuristic methods and numerical solvers like constraint programming-satisfiability (CP-SAT)—struggle to balance optimality and efficiency. Expert-driven routing is prohibitively time-consuming and prone to constraint violations. Heuristic methods, lacking a learning-based framework, often converge to local optima through repetitive iterations. Even robust numerical solvers face scalability bottlenecks, failing to provide satisfying solutions within reasonable computation time.

In another approach, reinforcement learning (RL)-based methodologies have been explored for global routing problems \citep{drla, drlb, drlc}. Early works demonstrated the feasibility of applying RL to global routing, showing performance advantages over heuristic approaches such as A* \citep{drla}. Subsequent studies introduced improved learning frameworks, such as double deep q-network (DDQN), to enhance routing performance and enable comparisons with both deep q-network (DQN) and heuristic methods \citep{drlb}. More recently, multi-agent reinforcement learning (MARL) has been adopted to incorporate local spatial information into state representations, allowing coordinated decision-making among multiple agents \citep{drlc}.
Despite these advances, RL-based approaches exhibit several limitations. Their performance is highly sensitive to parameter tuning, particularly the design of the reward function. Since global routing inherently involves both objective optimization and constraint satisfaction, the relative weighting between these components can significantly influence routing outcomes. Furthermore, selecting source/target pins (S/T pins) can yield different routing results even under identical configurations, highlighting the difficulty of capturing symmetric routing behavior within RL-based frameworks.

Recent studies have explored using generative artificial intelligence (Gen AI) to solve global routing problems. One representative approach employs a two-stage framework, where hubs required for net connections are first generated using Gen AI, followed by reinforcement learning-based optimization through the rectilinear edge sequence (RES) algorithm \citep{hubrouter}. Another approach introduces Diffusion Schrödinger Bridge (DSB) to enable end-to-end path generation without relying on a multi-stage process \citep{dsbrouter}. In addition, multi-agent-based frameworks have been proposed to incorporate information from other nets as additional inputs during the routing generation process \citep{glsvlsi}.
Despite these advances, several limitations remain. The two-stage framework requires additional computational resources to address connectivity issues inherent in generative models \citep{hubrouter}. 
The multi-agent approach focuses on refining partial routes rather than generating complete routing paths, and its effectiveness may be limited at early timesteps due to insufficient information in the self-attention mechanism \citep{glsvlsi}. Furthermore, DSB-based approaches require iterative gradient-based updates during inference, leading to increased computational time \citep{dsbrouter}. These limitations highlight the need for a generative routing framework that ensures connectivity while achieving faster and more accurate performance. 

This paper presents a hybrid global routing framework that integrates constrained A* with flow matching (FM). In the proposed approach, FM is employed to generate a global cost map that guides the constrained A* search process. Specifically, for each net-centric routing problem, a binary occupancy map is constructed by assigning target values indicating whether each node belongs to the routing path. The results are then used as training targets for FM to capture the underlying routing distribution. Subsequently, the cost map is obtained through ordinary differential equation (ODE)-based integration of the vector field and is utilized to guide the constrained A*-based routing.

Experimental results demonstrate that the proposed method achieves comparable total wirelength (WL) and via usage to baseline methods while consistently satisfying routing constraints. In addition, the framework is capable of generating solutions with reduced WL and via counts in certain instances, particularly for complex routing problems where conventional constrained A* fails or numerical solvers require excessive computation time. Furthermore, the proposed approach exhibits strong scalability, effectively handling problems with varying sizes and numbers of nets.

The main contributions are summarized as follows:

\begin{itemize}
    \item We propose a novel hybrid framework that integrates FM with constrained A*-based global routing, where FM is utilized to generate a global cost map that effectively guides the routing process.

    \item To the best of our knowledge, this is the first work that introduces a binary FM formulation for cost map generation, enabling the learning of node-level routing distributions based on path occupancy.

    \item The paper demonstrates a cost map based guidance generated by FM, including net ordering and dealing with heuristic functions. This suggests the further capability of using generated cost maps in other occasions.

    \item The proposed framework demonstrates improved scalability and search capability, consistently identifying feasible and high-quality routing solutions as problem complexity increases, including cases where conventional methods fail to find valid solutions or require significantly higher computational cost.
\end{itemize}

The rest of the paper is organized as follows. Section 2 explains preliminaries of global routing, along with problem definitions and algorithms used in this paper. The framework and settings in this study is described in Section 3. We argue about the results and discussions of the study in Section 4, and conclude the paper in section 5.

\section{Preliminaries}\label{sec2}

\subsection{Global Routing}\label{subsec2.1}

\begin{figure}
\centering
 \includegraphics[width=0.5\textwidth]{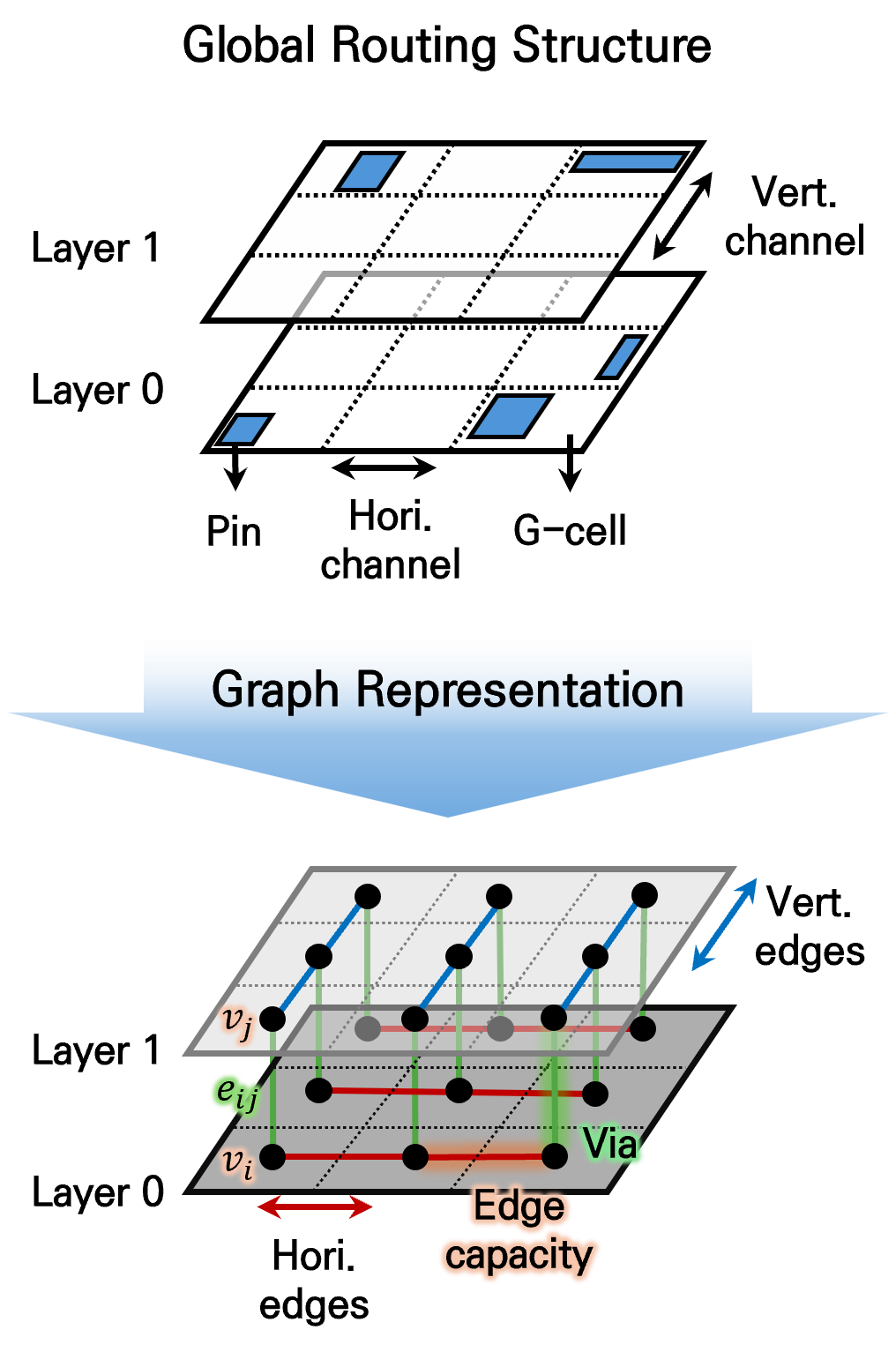}
\caption{Global routing structure and its graph representation.}
\label{fig:graph}
\vspace{-9pt}
\end{figure}

Global routing refers to the process of connecting pins on a given layout with predefined connections in units of grid cells containing pins, where the IC space can be considered as a grid space divided into many grid cells. Here, each grid cell is called a G-cell, and the grid space consisting of these G-cells can be modeled as a grid graph $G(V, E)$ as shown in Fig.~\ref{fig:graph}. Here, each vertex $v_i$ represents each G-cell divided in the chip space, and edge $e_{ij}$ represents a virtual line connecting $v_i$ and $v_j$. 
Due to the large number of pins to be connected, it is often infeasible to complete routing within a single metal layer. To handle this limitation, multiple routing layers are employed to provide independent paths and avoid routing conflicts. Inter-layer transitions are achieved by vias, vertical connections that link different metal layers.
In principle, the availability of sufficient multiple layers increases routing flexibility and allows all nets to be connected. However, vias introduce additional manufacturing costs and may degrade reliability compared to routing within a single layer. Therefore, an important objective in global routing is to minimize via usage while efficiently utilizing the available layer resources.

Each edge in the routing graph is associated with a capacity, which defines the maximum number of nets that can pass through it. An overflow (OF) occurs when the routing demand exceeds the capacity, leading to routing congestion and potential design-rule violations.
Therefore, the global routing optimization problem can be formulated as determining routing paths for all nets such that the total WL and via usage are minimized, while satisfying capacity constraints and avoiding routing conflicts. In this study, minimizing total WL and via usage is considered an objective function, whereas capacity constraints (to prevent OF) and net conflict avoidance are treated as constraints.

\subsection{Routing Algorithms}\label{subsec2.2}
\subsubsection{A* Algorithm}\label{subsubsec2.2.1}

The A* algorithm is one of the most widely used heuristic-based methods in IC global routing for finding a path between S/T pins. For each node $n$ in the grid graph, the algorithm evaluates a cost function that determines the search priority, defined as:

\begin{equation}
f(n) = g(n) + h(n)
\label{eq1}
\end{equation}

where $g(n)$ represents the cumulative wiring cost from the starting node to the current node $n$, and $h(n)$ denotes a heuristic function that is an estimate of the remaining distance to the target node. Note that $g(n)$ can include not only the simple geometric distance but also the congestion of each grid and the weights based on the use of vias. The Manhattan distance is generally used for calculating $h(n)$, and the lower bound is typically used to ensure optimality and efficiency at the same time.

The global routing problem considered in this study extends beyond minimizing total WL and requires satisfying various constraints imposed by manufacturing processes and design rules. To address these requirements, the constrained A* algorithm is adopted, which directly incorporates problem-specific constraints into the search process.
First, to enforce layer-dependent preferred routing directions, the search state is extended to a three-dimensional representation $(x, y, \text{layer})$. Based on the allowable routing direction of each layer, the expansion of neighboring nodes is dynamically restricted. In addition, via costs are incorporated into $g(n)$ to penalize unnecessary layer transitions and discourage excessive use of vias.
Next, to prevent routing conflicts between different nets, paths are restricted from passing through G-cells occupied by the terminals (S/T pins) of other nets. This is implemented by excluding such nodes from the open list during node expansion when they are already assigned to fixed pin regions of other nets. This constraint ensures physical separation between nets and helps maintain signal integrity.

\subsubsection{Rip-up and Reroute}\label{subsubsec2.2.2}

One of the key objectives of global routing is to ensure that all routing paths satisfy the edge capacity constraints of the grid graph. However, during the initial routing phase, each net independently searches for a shortest path, often resulting in multiple nets being assigned to the same edges. This leads to congestion, where routing demand exceeds available edge capacity, resulting in OF.

To handle this issue, RRR is employed as an iterative optimization technique. In RRR, nets contributing to OF are identified, their existing routes are removed (rip-up), and the cost of congested edges is increased to discourage repeated usage. The routing process is then repeated (re-routing), and heuristic algorithms such as A* use the updated cost function to search for alternative paths that avoid congested regions.

This iterative process continues until all nets are routed within the capacity constraints. In negotiated congestion-based routing frameworks such as PathFinder, RRR plays a central role by incorporating historical congestion information into the cost function, enabling adaptive coordination among competing nets and improving overall routability \citep{pathfinder}. However, routing approaches relying solely on RRR can exhibit an inherent limitation in that convergence is not guaranteed within a reasonable number of iterations, particularly for complex routing instances.

\subsubsection{Numerical Solvers}\label{subsubsec2.2.3}

Another powerful approach to addressing the global routing problem is to formulate it as a constraint satisfaction problem and solve it using a numerical optimization solver. In this study, we utilize the CP-SAT solver from Google OR-Tools, which integrates integer programming and constraint programming with the efficient search capabilities of a SAT solver~\citep{ortools}.

Conventional A*-based sequential routing connects nets one at a time and treats previously routed paths as obstacles, resulting in fast computation but exhibiting strong sensitivity to net ordering. This often leads to suboptimal routing, where later nets are forced to detour or fail to find feasible paths when earlier nets occupy critical resources. In contrast, solver-based approaches perform concurrent routing by modeling the paths of all nets as decision variables and optimizing them simultaneously. This enables global optimization over the entire design space and is particularly effective in resource-constrained environments where multiple nets compete for limited routing capacity.

A key advantage of the CP-SAT solver is its ability to efficiently prune infeasible regions during the search and identify solutions that minimize a weighted combination of WL and via usage. Compared to exhaustive search, this approach can handle significantly larger and more complex search spaces. Furthermore, solver-based methods help mitigate the limitations of heuristic approaches, which are prone to local optima, and serve as an effective tool for resource allocation in densely packed chip designs. On the other hand, the main drawback is the scalability issue, whereby the computation time required to resolve a problem increases as the problem scale grows.

\subsection{Flow Matching}\label{subsec2.3}

Flow Matching (FM) learns a continuous vector field $v_\theta$ that transforms the source noise distribution $p_i(z)$ into the target distribution $p(x)$, and the process of learning conditional relations between noise $e$ and the target $x_{gt}$ provides a tractable framework known as conditional flow matching (CFM). This defines the state $z_t$ at time $t \in [0, 1)$ as the linear interpolation between $e$ and $x_{gt}$. This can be expressed as follows~\citep{xpred}:

\begin{equation}
z_t = tx_{gt} + (1 - t)e, \quad e \sim \mathcal{N}(0, I).
\label{eq2}
\end{equation}

The corresponding conditional vector field $u_t$ is the derivative of $z_t$ with respect to time, and it is expressed as:

\begin{equation}
u_t(z_t | x_{gt}) = x_{gt} - e = \frac{x_{gt} - z_t}{1 - t}.
\label{eq3}
\end{equation}

This represents a straight path from the source to the destination at a constant speed, which aligns with the shortest distance between the two distributions---also known as the \textit{Optimal Transport} path---provided that $e$ and $x_{gt}$ are optimally coupled. This formulation simplifies the sampling process by minimizing the curvature of the trajectory during the generation process and leads to fewer errors in ODE-based integrations. Accordingly, high-quality samples can be achieved with a small number of steps. In addition, since this resulting target is constant, the model learns a clear and consistent direction, leading to faster convergence during training.

The generation process is formulated as an initial value problem, and the estimated signal $\hat{x}$ at $t = 1$ is acquired by integrating $v_{\theta}$, the velocity field obtained from the trained model. From an arbitrary start time $t_0 \in [0, 1)$, the process is expressed as:

\begin{equation}
\hat{x} = z_{t_0} + \int_{t_0}^{1} v_\theta(z_t, t) \, dt.
\label{eq4}
\end{equation}

Note that the trained model outputs the target value itself rather than the velocity in this study. Then, based on Eq.~\eqref{eq3}, the corresponding $v_{\theta}(z_t, t)$ is calculated from the model's output $x_{pred}$ for each timestep as follows:

\begin{equation}
v_{\theta}(z_t, t) = \frac{x_{pred} - z_t}{1 - t}.
\label{eq5}
\end{equation}

This allows the model to dynamically reset the best destination at each step and reliably reach the final data distribution by finding a straight path in that direction.

\section{Methodology}\label{sec3}
\subsection{Problem Definition}\label{subsec3.1}

\begin{algorithm}[t]
\caption{Hierarchical Data Generation for FM Training}
\label{alg:data_generation}
\begin{algorithmic}[1]
\Require Routing instance $\mathcal{P}$, expansion limit $N_{\max}$, maximum RRR iterations $K_{\max}$, solver time limit $T_{\max}$
\Ensure Feasible routing solution $\mathcal{R}$

\State $\mathcal{R} \gets \textsc{Constrained A*}(\mathcal{P}, N_{\max})$
\If{$\mathcal{R}$ is feasible}
    \State \Return $\mathcal{R}$
\EndIf

\For{$k = 1$ to $K_{\max}$}
    \State Update routing costs based on congestion RRR
    \State $\mathcal{R} \gets \textsc{Constrained A*}(\mathcal{P}, N_{\max})$
    \If{$\mathcal{R}$ is feasible}
        \State \Return $\mathcal{R}$
    \EndIf
\EndFor

\State $\mathcal{R} \gets \textsc{CPSATSolver}(\mathcal{P}, T_{\max})$
\State \Return feasible solution that minimizes the objective
\end{algorithmic}
\end{algorithm}

In this study, training data for FM is generated using a combination of constrained A* and the CP-SAT-based solver from Google OR-Tools \citep{ortools}. The data generation process follows a hierarchical strategy to efficiently obtain feasible routing solutions under various constraints, as outlined in Algorithm~\ref{alg:data_generation}.

First, each routing instance is solved using a constrained A* algorithm with a predefined maximum number of node expansions for each net. If a feasible solution satisfying all constraints is not found within this limit, the RRR procedure is applied for up to 10 iterations to resolve congestion and improve routability, where the routing order is given in ascending order of Manhattan distance between the S/T pins of each net. For instances where feasible solutions are still not obtained, a CP-SAT-based solver is employed as a fallback mechanism. The solver is executed with a time limit of 180 seconds, and the best feasible solution found within this limit, in terms of the defined objectives, is selected.

All nets are modeled as 2-pin connections, and each G-cell is assumed to contain a single pin. To prevent short-circuit issues between different nets \citep{mlshort, ehpred, shortfinder, routenet, vionet}, a constraint is imposed such that routing paths of other nets are prohibited from passing through the S/T pins of a given net.
Accordingly, the routing constraints considered in this study include edge capacity violations (OF) and S/T pin conflicts. Additional problem settings are summarized in Table~\ref{tab1}.
To evaluate the generality of the proposed method, the model is trained on routing instances with varying problem sizes and numbers of nets within a predefined range.

\renewcommand{\arraystretch}{1.5}
\begin{table}[b]
\centering
\caption{Problem settings for training.}\label{tab1}%
\begin{tabular}{@{}cclll@{}}
\toprule
Settings & \multicolumn{4}{c}{Specifications} \\  \midrule
Initial edge capacity (x, y, z) & \multicolumn{4}{c}{(1, 1, 1)} \\
Data size (constrained A*/CP-SAT) & \multicolumn{4}{c}{2500/2500} \\
Max expansion size per net & \multicolumn{4}{c}{50,000} \\
Train/test division ratio & \multicolumn{4}{c}{8:2} \\
Grid size per data (x, y, z) & \multicolumn{4}{c}{(25, 25, 2) - (30, 30, 2)} \\
Number of nets per data & \multicolumn{4}{c}{15 - 20} \\
Time limit for CP-SAT (sec) & \multicolumn{4}{c}{180} \\ \bottomrule
\end{tabular}
\end{table}

The generated routing solutions are decomposed into net-wise representations for preprocessing. This enables the FM-based model to learn cost maps at the level of individual nets while supporting batch processing across all nets within a given routing instance.
For each net, a binary occupancy map is constructed, where each node is assigned a value of 1 if it belongs to the routing path of the corresponding net and 0 otherwise. In addition, two auxiliary maps are defined: a current pin map indicating the locations of the S/T pins of the target net, and an obstacle map representing the S/T pin locations of the other nets in the same instance. These are later used as input to the network, providing guidance on which pins to avoid and which to connect for each net.

\subsection{Framework}\label{subsec3.2}
\begin{figure}
\centering
 \includegraphics[width=1\textwidth]{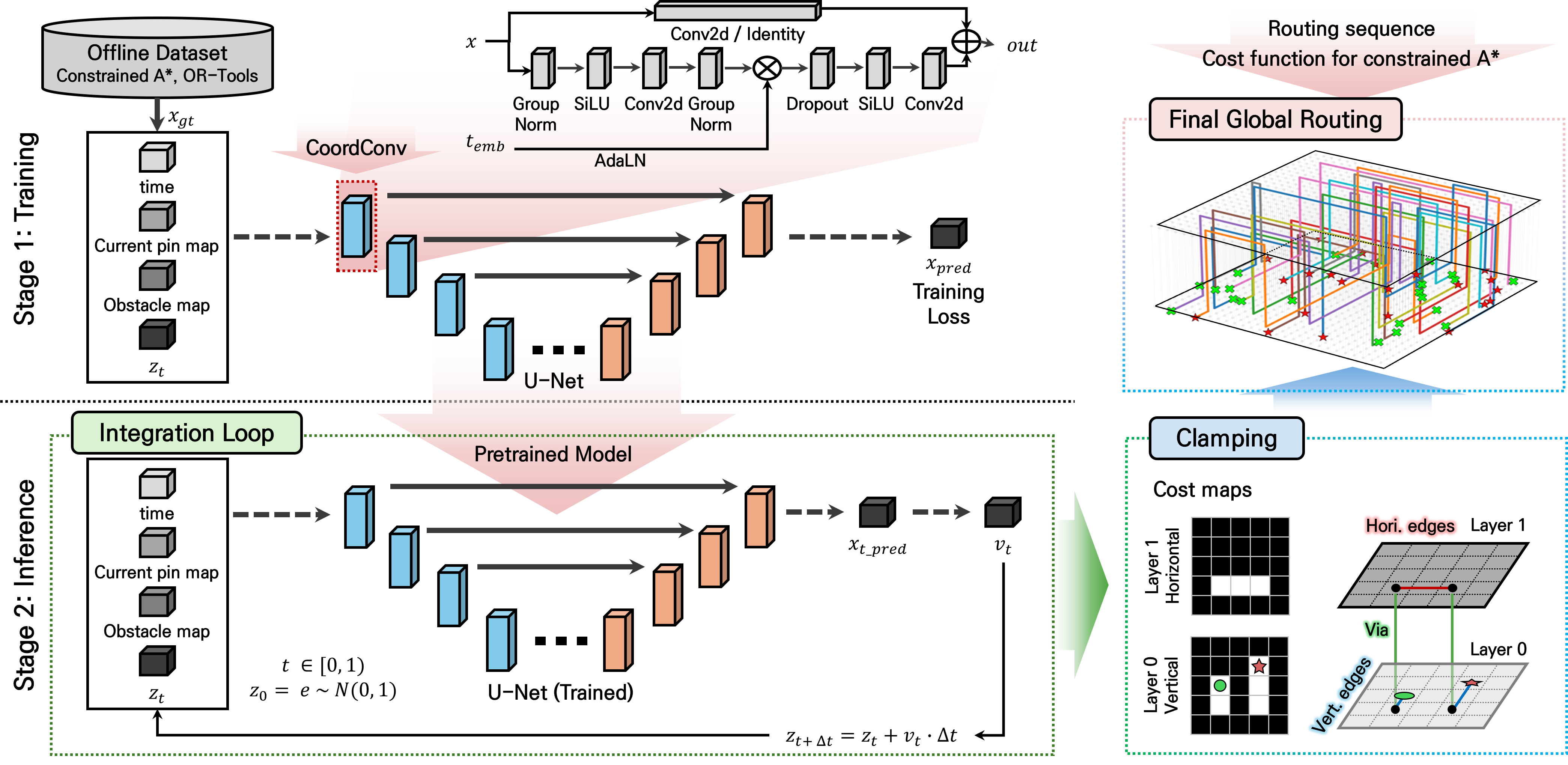}
\caption{Framework of the proposed method.}
\label{fig:framework}
\vspace{-9pt}
\end{figure}

The proposed framework consists of two main stages, namely a training stage and an inference stage, as illustrated in Fig.~\ref{fig:framework}. The overall objective of the framework is to learn a cost map representation that captures global routing tendencies and to utilize this representation to guide a constrained A*-based routing process.

In Stage 1: training (Section~\ref{subsubsec3.2.1}), a neural network is trained via FM to generate cost maps for individual nets based on routing instances. The training data are generated using a combination of constrained A*, RRR, and a CP-SAT-based solver, enabling the model to learn from feasible routing solutions under various constraints.

In Stage 2: inference (Section~\ref{subsubsec3.2.2}), the trained model generates cost maps for unseen routing instances. These cost maps are incorporated into the evaluation function of the constrained A* algorithm, guiding the search toward feasible, low-cost routing paths. This hybrid strategy combines the global guidance capability of the generative model with the feasibility guarantees of heuristic search.
The following subsections describe each stage of the framework in detail:

\subsubsection{Stage 1: Training}\label{subsubsec3.2.1}

Stage 1 focuses on training a neural network to generate cost maps based on FM. A U-Net-based architecture is adopted for this purpose, as it is widely used in diffusion-based models \citep{unet}.
Specifically, the network is composed of multiple blocks that incorporate adaptive layer normalization (AdaLN) and ResNet modules \citep{adaln, resnet}. Within each block, skip connections are implemented using either identity mappings or $2$D convolutional layers, depending on whether channel alignment is required.

For the input, layer-by-layer images are constructed by separating the routing layers of each chip into individual channels. For each net, the input consists of the time variable $t$ and the corresponding state $z_t$ as defined in \eqref{eq2}, along with the current pin map and obstacle map.
The same time value is provided for all nets within a given chip to ensure consistency, as the network operates on a per-net basis while ultimately generating routing solutions jointly across all nets. Additionally, CoordConv channels are incorporated in the first block to enhance the learning of absolute spatial coordinates \citep{coordconv}.

Unlike conventional FM, the proposed method adopts an $x$-prediction with $x$-loss formulation tailored for binary FM training, enabling the generation of cost maps aligned with binary targets for each net \citep{xpred}. Under this formulation, the model is trained to directly predict the target value at each node (pixel), rather than the velocity field, and the loss is computed based on the discrepancy between the predicted output and the target.

To implement the analog-bit input/output design, the current pin map and obstacle map are rescaled to the range $[-1, 1]$, and the target values are transformed using a logit function \citep{xpred}. Let the original target value be $x_{\mathrm{tgt}} \in [0, 1]$ and the transformed value be $x_{\mathrm{tgt}, t}$; the transformation is defined as follows, where $\epsilon = 0.05$.

\begin{equation}
x_{tgt, t} = \log \left( \frac{\text{clamp}(x_{tgt}, \epsilon, 1 - \epsilon)}{1 - \text{clamp}(x_{tgt}, \epsilon, 1 - \epsilon)} \right)
\label{eq6}
\end{equation}
\\

The training loss consists of an MSE-based mask loss designed to ensure the model follows the target~\citep{glsvlsi}. The function $\mathbb{I}$ in Eq.~\eqref{eq8} refers to the indicator function. Note that padding is required for grouping data with different shapes into batches for U-Net; therefore, the pixels corresponding to the padding were excluded when calculating the loss, and batch-level dynamic padding was used to minimize the padding. A detailed explanation of the specific parameters is shown in Table~\ref{tab:training}.

\begin{equation}
\mathcal{L}_{m} = \frac{1}{N_{v}} \sum_{i} \Big[ \alpha_{m} M_i + \beta_{m} (1 - M_i) \Big](l_i - y_i)^2
\label{eq7}
\end{equation}

\begin{equation}
M_i = 
\begin{cases} 
1, & \text{if } \mathbb{I}(y_i > 0) + \mathbb{I}(\text{sigmoid}(l_i) > 0.5) \ge 1, \\
0, & \text{otherwise,}
\end{cases}
\label{eq8}
\end{equation}

\begin{equation}
O_i = 
\begin{cases} 
1, & \text{if } i \in \mathcal{O}, \\
0, & \text{otherwise,}
\end{cases}
\label{eq9}
\end{equation}

\renewcommand{\arraystretch}{1.5}
\begin{table}[t]
\centering
\caption{Parameters for training stage.}\label{tab:training}
\begin{tabular}{@{}ccc@{}}
\toprule
Parameter & Meaning & Value \\ \midrule
$N_{v}$ & Number of valid pixels & - \\
$\alpha_{m}$ & Penalty weight for $M_i$ & 2 \\
$\beta_{m}$ & Penalty weight for $1 - M_i$ & 1 \\
$l_i$ & Raw model output for pixel $i$ & - \\
$y_i$ & $x_{tgt, t}$ for pixel $i$ & - \\
$\mathcal{O}$ & Set of pixels included in obstacle map & - \\ \bottomrule
\end{tabular}
\end{table}

\subsubsection{Stage 2: Inference}\label{subsubsec3.2.2}

Stage 2 utilizes the trained FM model to generate a cost map for a given routing instance, which is then used to guide the constrained A*-based routing process (as shown in Fig.~\ref{fig:framework}). 
The cost map is obtained by iteratively integrating the ODE defined in Eq.~\eqref{eq5}. In this study, the Heun2 method is employed with 20 discretized time steps. Compared to conventional diffusion-based approaches, FM is capable of achieving high-quality results with a significantly smaller number of integration steps, thereby reducing inference time.
A detailed analysis of the cost maps generated with only 20 steps is provided in Section~\ref{sec4}. The resulting cost map is further processed via pseudo-binarization, where values below 0.5 are set to 0 and those above are retained.

While the FM model generates cost maps for each net independently, using these maps directly for routing may result in ambiguous or conflicting regions across nets. In particular, high-probability overlapping regions indicate potential competition for shared routing resources, which can result in capacity violations such as OF.
To address these challenges, the cost maps are not used as direct routing solutions but instead serve as guidance for the constrained A*-based routing process. By incorporating the generated cost maps into the evaluation function of A*, the routing process is biased toward regions with high likelihood as described in the cost maps, while preserving the feasibility guarantees of the heuristic search.
This hybrid strategy effectively combines the strengths of both approaches: the generative model provides global guidance over the solution space, while the constrained A* ensures routability and strict satisfaction of constraints. As a result, the framework overcomes the inherent limitation of purely generative methods in ensuring feasible routing solutions.

\begin{figure}
\centering
 \includegraphics[width=0.5\textwidth]{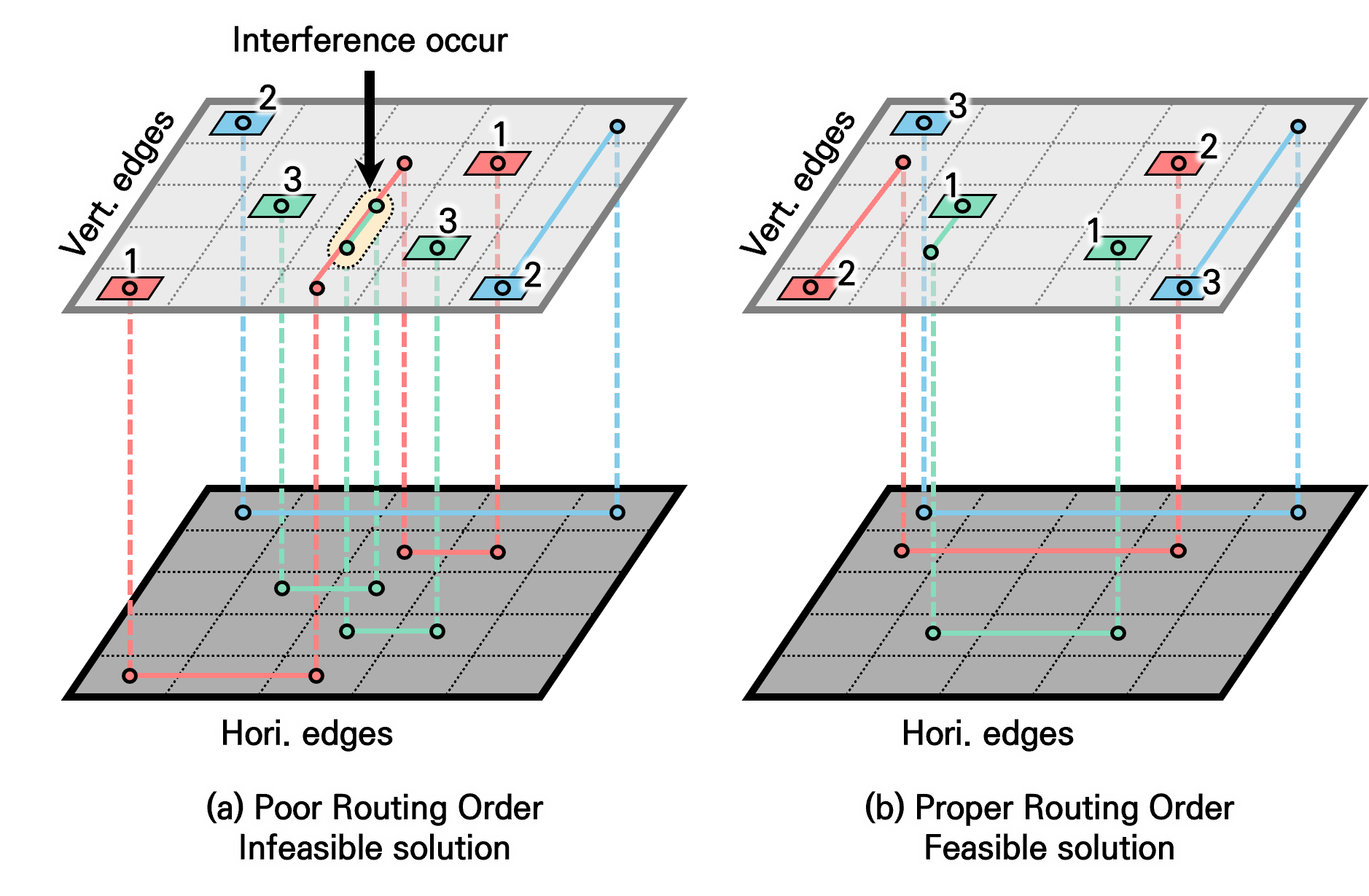}
\caption{Effect of routing order in sequential A*-based routing. Even with identical pin configurations, different routing sequences can yield feasible or infeasible solutions due to resource conflicts.}
\label{fig:Routingorder}
\vspace{-9pt}
\end{figure}

As mentioned before, the A*-based routing methodology relies on sequential routing, where nets are processed one at a time. As a result, the routing order significantly influences the feasibility of the final solution, and improper ordering can lead to routing failure even when feasible solutions exist, as shown in Fig.~\ref{fig:Routingorder}. This issue has been widely recognized and studied in prior works \citep{comm, drcu, tao, async, trans}.
In practice, determining an optimal routing order is closely aligned with human intuition in manual routing, where initial routing decisions critically affect the routability of subsequent nets. However, such ordering strategies are inherently suboptimal and lack robustness, especially in complex and highly congested scenarios.
Furthermore, conventional A* typically uses the Manhattan distance as the heuristic $h(n)$ to estimate the remaining cost. While computationally efficient, this heuristic does not account for congestion or resource competition across the routing space. As a result, it provides only coarse directional guidance and fails to capture the global structure of feasible routing solutions. Thus, Manhattan distance alone cannot account for the varying levels of congestion, and congestion itself can only provide a general direction for routing; it cannot specify the exact path each net should take, which is a critical limitation.

This study proposes a cost-guided routing strategy that jointly considers routing order and heuristic design for constrained A*, thereby improving routing efficiency and robustness. The key idea is to leverage the cost maps generated for all nets to both prioritize routing order and guide the search process. For routing-order determination, the cost maps of all nets are first computed simultaneously. Based on these maps, a priority metric is defined for each net as the ratio of pixels that spatially overlap with other nets relative to its total path length. Nets with higher overlap ratios are routed earlier, as they are more likely to occupy highly congested regions and compete for limited routing resources. By prioritizing such nets, routing decisions in critical regions are decided before resources are exhausted, while nets located in less congested areas are deferred to later stages. This global ordering strategy can improve overall routing efficiency compared to conventional approaches such as the area of the region covered by a net. However, routing order alone may be insufficient to fully exploit the information contained in the cost maps. To effectively utilize the guidance, the cost map is incorporated directly into the evaluation function of constrained A*. 

Specifically, the conventional A* evaluation function in Eq.~\eqref{eq1} is modified to incorporate cost map guidance, as defined in Eqs.~\eqref{eq10}--\eqref{eq12}, enabling more informed exploration of the routing space.

\begin{equation}
h_{m}(n) = |x| + |y| + C_z|z|
\label{eq10}
\end{equation}

\begin{equation}
f(n) = g(n) + h'(n)
\label{eq11}
\end{equation}

\begin{equation}
h'(n) = h_{m}(n) + \alpha (1 - P_{own}(n)) + \beta P_{other}(n).
\label{eq12}
\end{equation}

Here, the heuristic term $h_{m}(n)$ in Eq.~\eqref{eq10} denotes the Manhattan-based heuristic extended to three-dimensional routing, where $C_z$ represents the additional cost associated with inter-layer (via) transitions and is set to 5 in this study. The terms $P_{own}(n)$ and $P_{other}(n)$ in Eq.~\eqref{eq12} correspond to the cost map outputs of the FM model, representing the likelihood of the occupying node $n$ of the current net and the likelihood of occupation by other nets or obstacles, respectively. Note that $P_{other}(n)$ is a normalized value of the sum of the occupation likelihoods of the other nets.

As defined in Eq.~\eqref{eq12}, the node traversal cost is adjusted to encourage routing through regions with high confidence for the current net while discouraging paths that may lead to conflicts with other nets. Specifically, $1 - P_{own}(n)$ reduces the effective cost, whereas $P_{other}(n)$ increases it to account for potential congestion. The final evaluation function in Eq.~\eqref{eq11} combines these terms to guide the search toward feasible and cost-efficient routing paths.

The weighting parameters $\alpha$ and $\beta$ in Eq.~\eqref{eq12} are dynamically decayed as the search progresses to prevent the cost map from overly biasing the routing process. In particular, both weights gradually decrease as the number of node expansions increases, allowing alternative path explorations when the initial guidance becomes unreliable. The decay formulation is given as follows, where $\alpha_0$, $\beta_0$, $D_{th}$, $\mathrm{cnt}$, and $\mathrm{cnt}_{\max}$ denote the initial weights, decay threshold, and the current and maximum number of node expansions per net, respectively. The main parameter settings are summarized in Table~\ref{tab:inference}.

\begin{equation}
\alpha = \alpha_0 \cdot \text{max}\left(0, 1 - \frac{\text{cnt}}{D_{th} \cdot \text{cnt}_{max}}\right)
\label{eq13}
\end{equation}

\begin{equation}
\beta = \beta_0 \cdot \text{max}\left(0, 1 - \frac{\text{cnt}}{D_{th} \cdot \text{cnt}_{max}}\right).
\label{eq14}
\end{equation}

\renewcommand{\arraystretch}{1.5}
\begin{table}[h]
\centering
\caption{Parameters for inference stage.}\label{tab:inference}
\begin{tabular}{@{}ccc@{}}
\toprule
Parameter & Meaning & Value \\ \midrule
$\alpha_0$ & Initial coefficient for $P_{own}$ & 2 \\
$\beta_0$ & Initial coefficient for $P_{other}$ & 1.5 \\
$D_{th}$ & Decay ratio & 0.5 \\
$\text{cnt}_{max}$ & Maximum expansion sizes per net & 50,000 \\
$max_{rrr}$ & Maximum RRR & 10 \\
\bottomrule
\end{tabular}
\end{table}

The incorporation of the FM-based cost map modifies the heuristic function $h'(n)$ in a way that may violate the strict admissibility condition of the conventional A* algorithm. This means that the additional cost terms can lead to an overestimation of the true cost-to-go.
However, this behavior is analogous to Weighted A*, where the heuristic is intentionally biased to accelerate the search. By
reshaping the cost landscape, the proposed heuristic encourages exploration to focus on regions more likely to yield feasible, high-quality solutions, as indicated by the cost map. As a result, the search process becomes more efficient, significantly reducing the number of expanded nodes while maintaining near-optimal routing performance.
In practice, this leads to bounded suboptimality, where the resulting routing cost deviates only marginally from the optimal solution, offering a favorable trade-off between solution quality and computational efficiency for complex global routing problems.

\begin{figure}
\centering
 \includegraphics[width=0.5\textwidth]{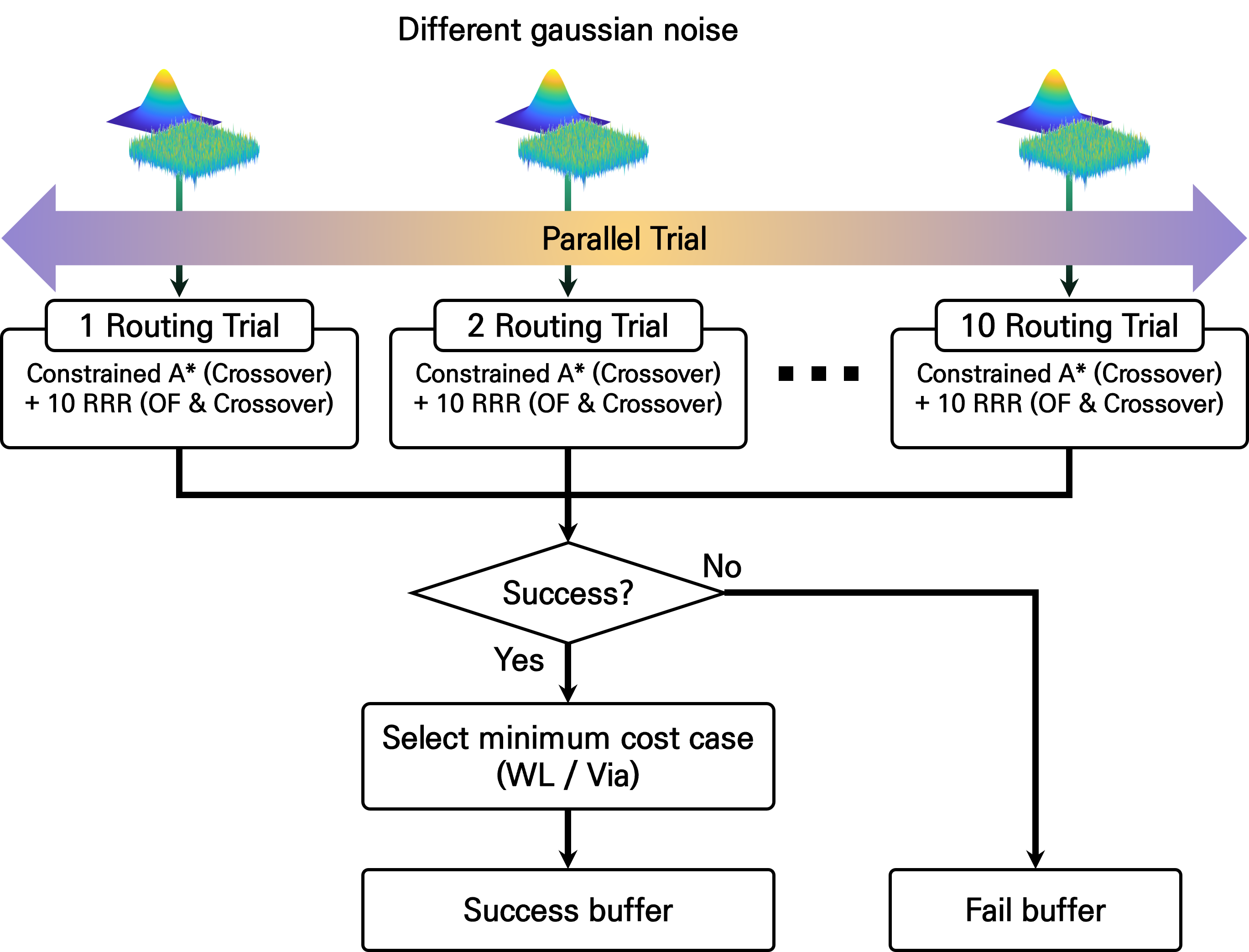}
\caption{Process of inference stage.}
\label{fig:method}
\vspace{-9pt}
\end{figure}

In the inference stage, the final routing solution is obtained by applying constrained A* with the modified heuristic, followed by up to 10 iterations of RRR, consistent with the data generation process. Furthermore, to account for stochastic variations in the generated cost maps, $K=10$ independent noise samples are drawn for each routing instance, and the best solution according to the objective function is selected. The overall inference flow is illustrated in Fig.~\ref{fig:method}.

\section{Experimental Results}\label{sec4}
This section presents experimental results to evaluate the effectiveness of the proposed framework. First, Section~\ref{subsec4.1} analyzes the training performance of the FM-based model by examining the generated cost maps. Next, Section~\ref{subsec4.2} investigates the impact of key design components through an ablation study, including the cost-guided heuristic and routing order strategy. Finally, Section~\ref{subsec4.3} demonstrates the practical advantages of the proposed method in complex routing scenarios through case studies.

\subsection{Training Results}\label{subsec4.1}

\begin{figure}
\centering
 \includegraphics[width=0.87\textwidth]{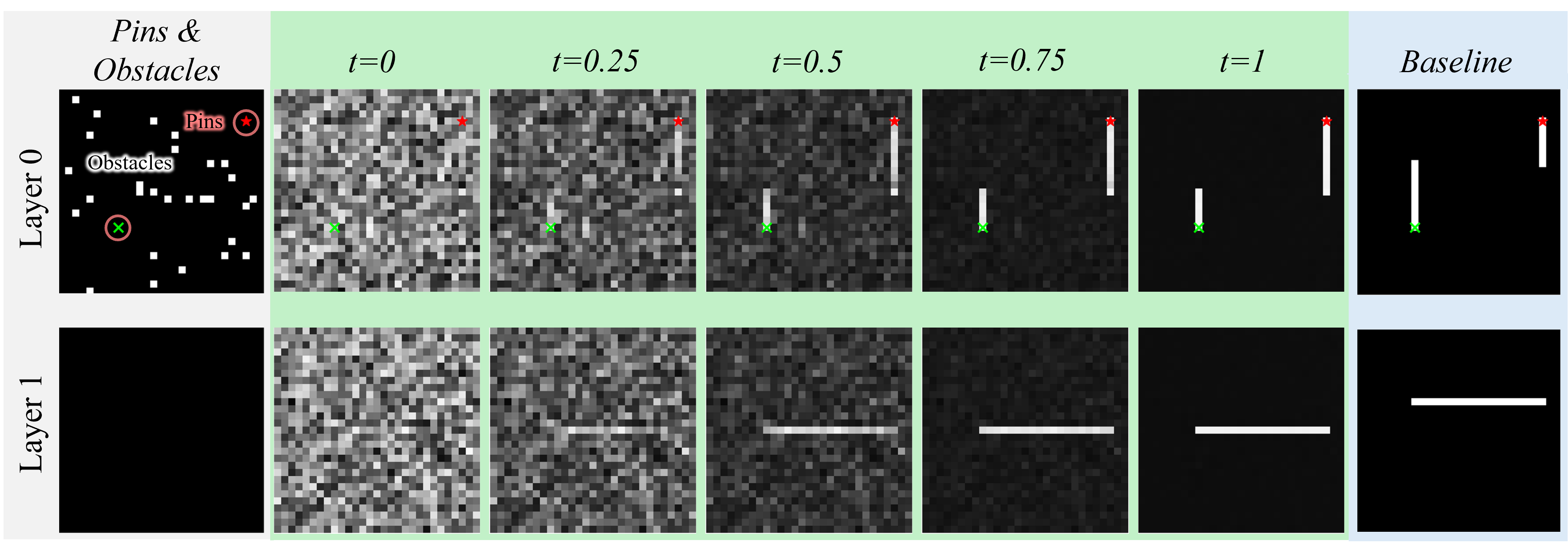}
\caption{Generation result of the cost map.}
\label{fig:generation}
\vspace{-9pt}
\end{figure}

Training was performed using the dataset described in Section~\ref{subsec3.1}, and the detailed training specifications are summarized in Table~\ref{tab4}. After training, the quality of the generated cost maps was evaluated, as shown in Fig.~\ref{fig:generation}. In the figure, the leftmost panel shows the S/T pin locations of the target net and the obstacle map constructed from the S/T pins of other nets within the same routing instance. The subsequent panels illustrate the cost map generation process for $t \in \{0, 0.25, 0.5, 0.75, 1\}$, while the rightmost panel shows the baseline result. Each row corresponds to different routing layers (layer 0 and layer 1, from top to bottom). From left ($t=0$) to right ($t=1$), the results demonstrate that the model progressively refines the cost map by suppressing noise over time, resulting in stable and coherent routing guidance.

\renewcommand{\arraystretch}{1.5}
\begin{table}[h]
\centering
\caption{Parameters for network training.}\label{tab4}
\begin{tabular}{@{}ccc@{}}
\toprule
Parameter & Meaning & Method/Value \\ \midrule
$opt$ & Optimizer & AdamW \\
$sch$ & Scheduler & Cosine Annealing \\
$lr$ & Learning rate & $10^{-4} -  10^{-5}$ \\
$wd$ & Weight decay & $10^{-4}$ \\
$batch$ & Batch size & 64 \\
$epoch$ & Training epochs & 50 \\
\bottomrule
\end{tabular}
\end{table}

\begin{figure}
\centering
 \includegraphics[width=0.87\textwidth]{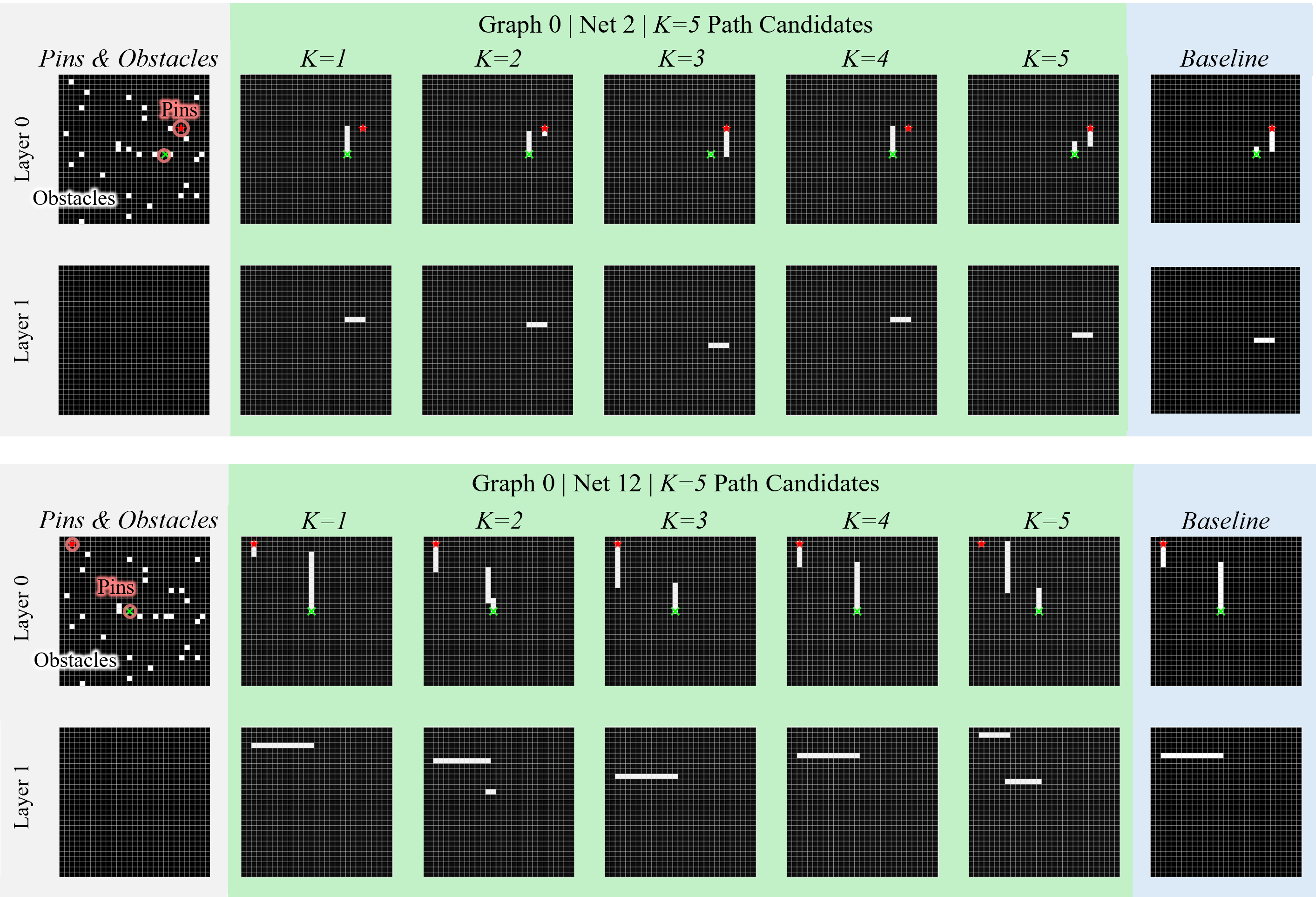}
\caption{Output results from 5 different noise sources.}
\label{fig:noise}
\vspace{-9pt}
\end{figure}

Due to the stochastic nature of FM, the generated cost maps may vary with input noise, which can influence the final routing outcome. To assess this variability, the cost maps generated from multiple noise samples are analyzed.
Specifically, five independent noise inputs are used, and the corresponding results are presented in Fig.~\ref{fig:noise}.
The overall structure of the figure is the same as in Fig.~\ref{fig:generation}, except the middle part (from the 2nd to 6th column) showing the generation results from 5 different noise sample inputs.

\begin{figure}
\centering
 \includegraphics[width=0.5\textwidth]{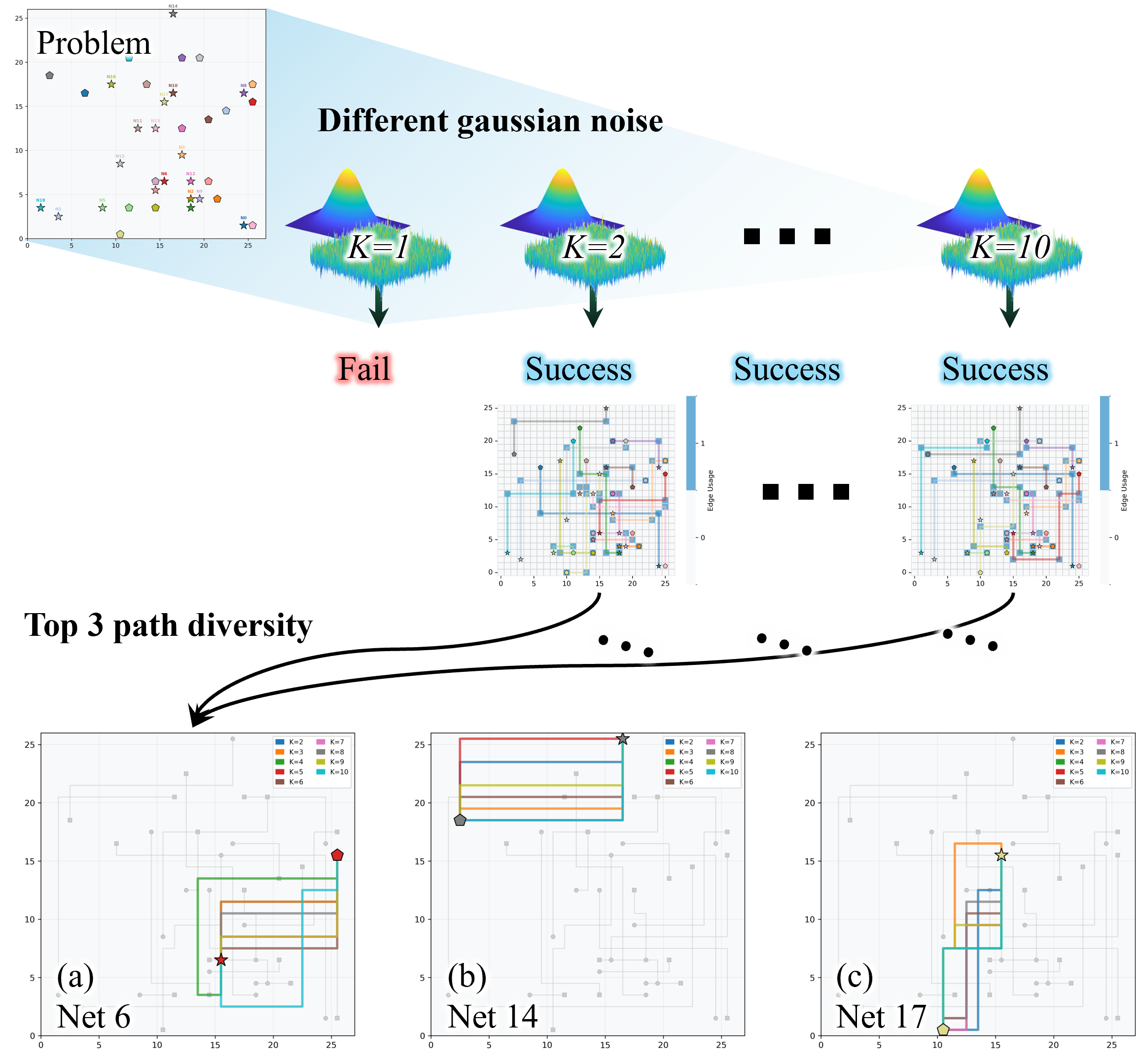}
\caption{Multiple suboptimal solutions generated by the FM-guided router.
        (a)--(c): Path diversity of selected nets across $K$ trials
        (colored lines: paths of the selected net with different $K$, gray: paths of the other nets with the best result).}
\label{fig:generative results}
\vspace{-9pt}
\end{figure}

Several observations can be made from Fig.~\ref{fig:noise}. First, the cost maps generated from different noise samples exhibit noticeable variability, indicating the stochastic nature of the FM model. Despite these differences, all generated maps consistently connect the S/T pins of the target net. This suggests that the model has not merely memorized the target $x_{tgt}$, but has learned its underlying distribution and can generate structurally valid cost maps under varying conditions.
Furthermore, the generated maps remain clean and coherent across different noise samples while satisfying layer-wise routing constraints. This means that the model captures both the spatial structure of feasible routing solutions and the associated constraints, enabling stable generation despite stochastic perturbations.

To further evaluate the influence of this variability, the final routing results obtained from different cost maps are compared for the same routing instance, where the routing results are demonstrated in a bird’s-eye view. As shown in Fig.~\ref{fig:generative results}, the resulting routes vary depending on the sampled cost maps. Based on this observation, the solution with the minimum WL and via usage is selected as the final result, although alternative selection criteria may be applied depending on the application.

\begin{figure}
\centering
 \includegraphics[width=0.7\textwidth]{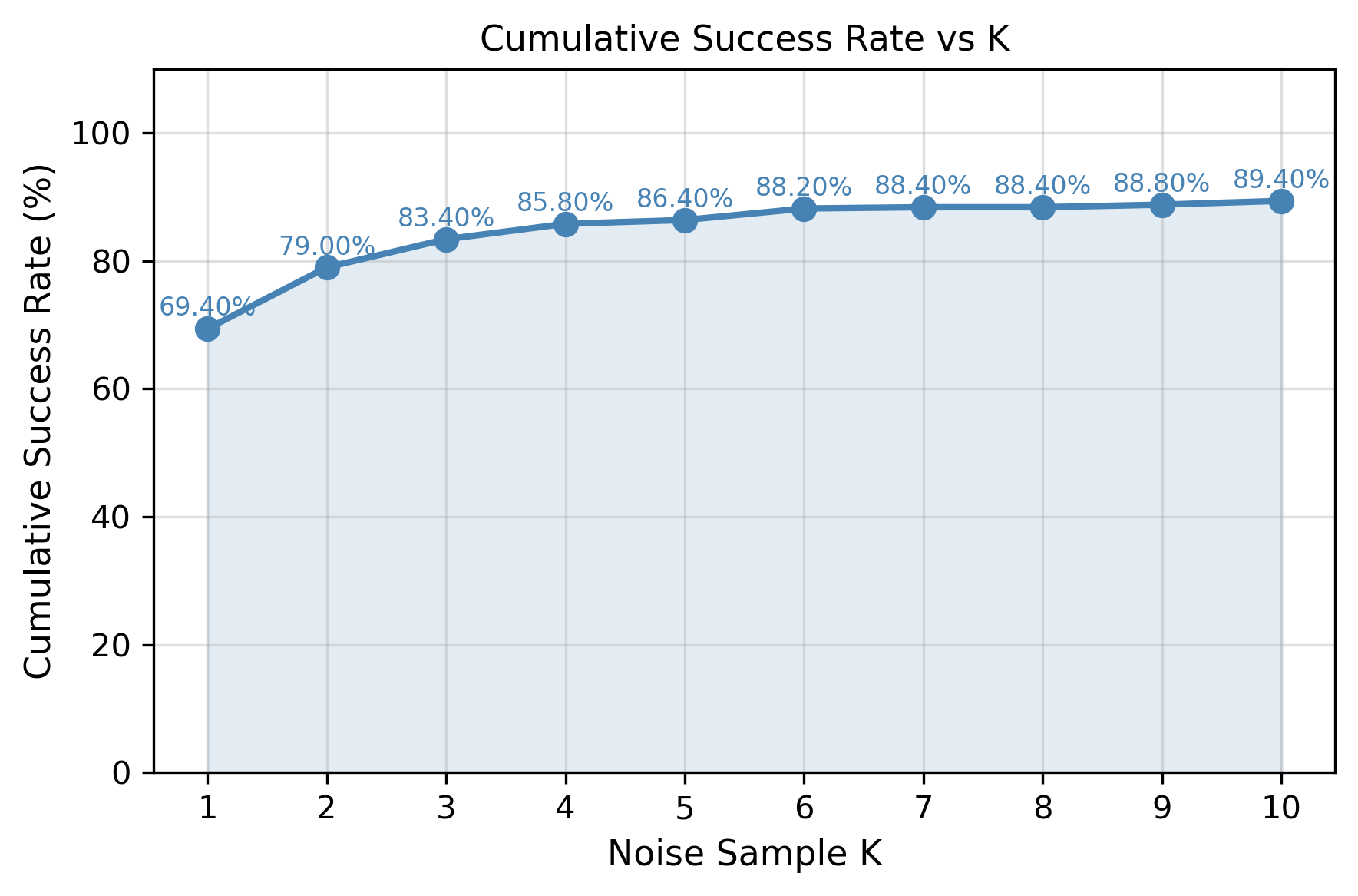}
\caption{Trend in Cumulative Success Rate by different noise samples.}
\label{fig:cumulative}
\vspace{-9pt}
\end{figure}

In addition, Fig.~\ref{fig:cumulative} presents the incremental success rate as the number of noise samples $K$ increases. The results show that the success rate improves with larger $K$, while also converging around $K=10$. This indicates that sampling multiple cost maps increases the likelihood of obtaining a feasible, high-quality routing solution, and 10 different trials can be a suitable option for cost map generation.

\clearpage

\subsection{Ablation Study}\label{subsec4.2}

An ablation study is conducted to evaluate the effectiveness of the two key components introduced in Section~\ref{subsubsec3.2.2}, namely the cost map-based routing order strategy and the modified heuristic function. Specifically, the final routing performance is compared under different configurations, depending on whether the cost map is used for routing order determination, heuristic modification, or both.
As a baseline, the case where neither component is applied corresponds to the constrained A*-based routing used for generating the initial $x_{tgt}$. This allows for a direct assessment of the contribution of each component.
To ensure robustness, all experiments are conducted using five different random seeds, and the results are averaged. The quantitative results are summarized in Table~\ref{tab5}. The WL metric includes via transitions, where each layer change is counted as a unit length.

The results in Table~\ref{tab5} demonstrate that the best overall performance is achieved when both the cost-map-based routing order and the heuristic modification are applied. In particular, the proposed method consistently achieves lower WL than the baseline across all datasets.
In terms of routing success rate, the method attains near-perfect performance on instances solvable by the pure constrained A* approach, while achieving approximately 90\% success rate on more challenging instances where the baseline fails. This indicates that the proposed framework provides effective guidance for exploring feasible routing solutions beyond the capability of conventional heuristics.
Regarding computational efficiency, the proposed method requires approximately 3 seconds per graph, which is less than 3\% of the runtime of solver-based approaches. This demonstrates that, even when accounting for network inference and ODE integration, the method offers a significant speed advantage over traditional optimization-based methods.

\begin{figure}
\centering
 \includegraphics[width=0.8\textwidth]{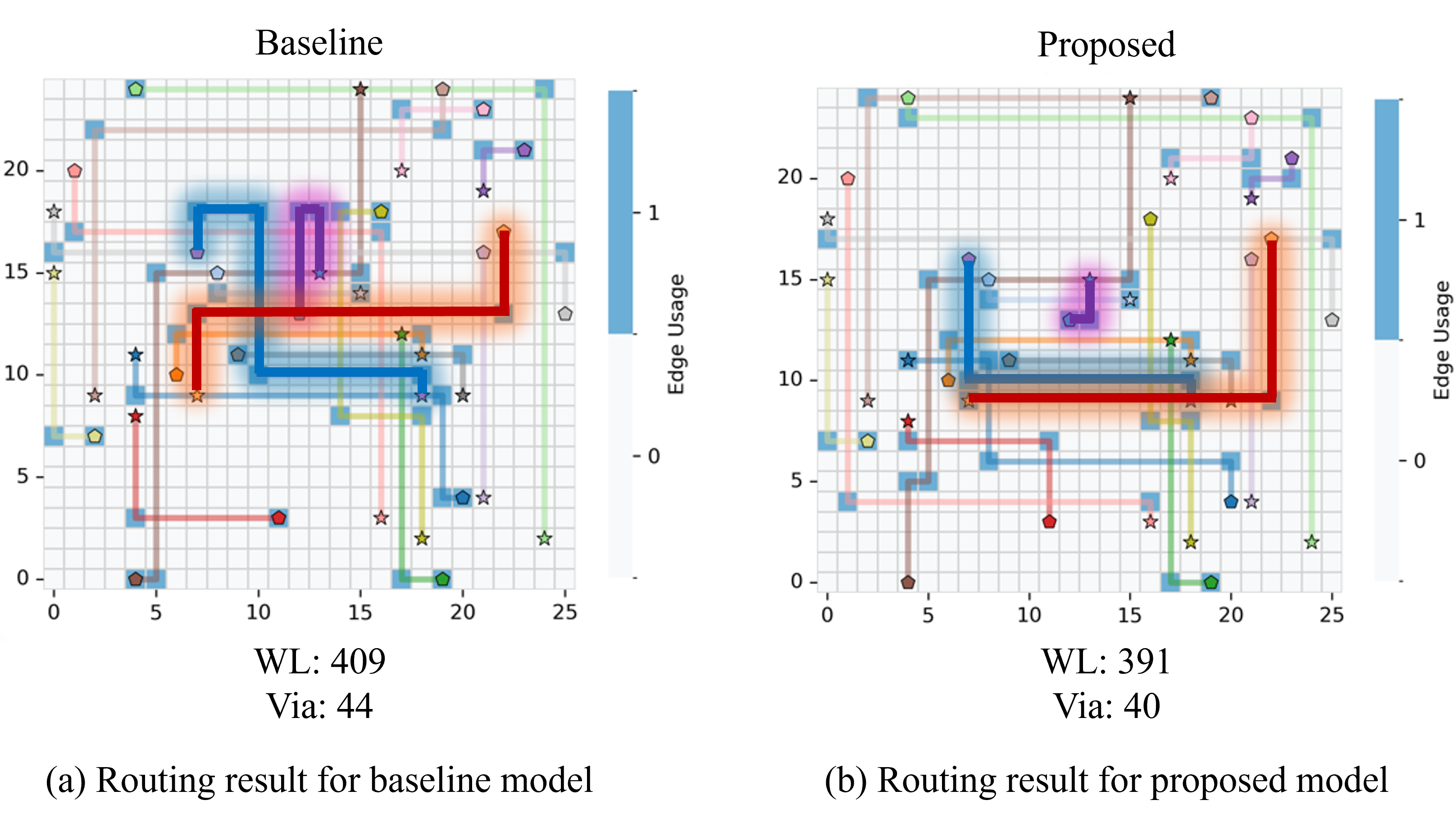}
\caption{Routing results comparison.}
\label{fig:Success comparison}
\vspace{-9pt}
\end{figure}

To further explain these improvements, representative routing examples are shown in Fig.~\ref{fig:Success comparison}. For identical S/T pin configurations, the proposed method produces routes with reduced WL and fewer vias compared to the baseline. In particular, in cross-routing grid environments, changes in routing direction often require inter-layer transitions, resulting in increased path bends. The proposed method mitigates this effect by generating smoother routes with fewer layer transitions, resulting in more efficient routing structures.

\begin{figure}
\centering
 \includegraphics[width=0.8\textwidth]{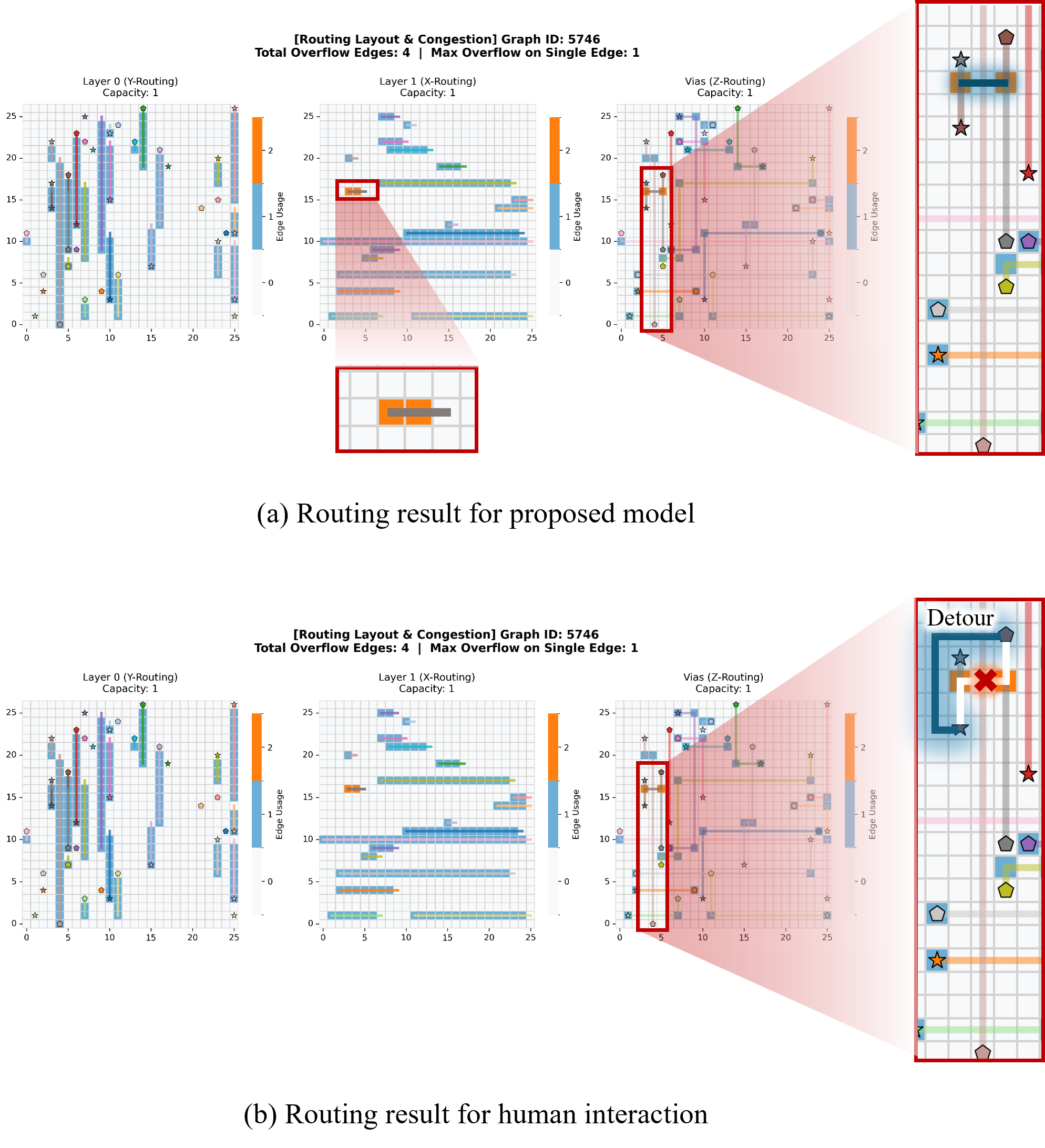}
\caption{Routing results modification.}
\label{fig:Failure comparison}
\vspace{-9pt}
\end{figure}

In addition, failure cases are analyzed to better understand the limitations of the proposed method. A representative example is shown in Fig.~\ref{fig:Failure comparison}, where the utilization of layers 0 and 1, and vias is visualized at the node level. For vias, both node-wise usage and the overall routing structure are presented.
The analysis reveals that routing failures are primarily caused by localized OF regions, rather than global infeasibility. As illustrated in the figure, minor modifications in these congested regions can resolve the conflicts and enable all nets to be successfully routed without OF.
This observation suggests that the proposed method produces near-feasible solutions even in failure cases, indicating strong potential for further improvement through local refinement or post-processing.

\clearpage

\renewcommand{\arraystretch}{1.3}
\begin{table}[t]
    \centering
    \caption{Ablation study results and runtime comparison.}
    \label{tab5}

    \begin{tabular}{llcccc}
    \toprule
    \multirow{2}{*}{Metric} & \multirow{2}{*}{Scenario}
        & \multicolumn{4}{c}{Method} \\
    \cmidrule(l){3-6}
     & & FMR+FMH & FMR+MH & MR+FMH & MR+MH \\
    \midrule
    \multirow{4}{*}{Success Rate (\%)}
     & Tr A* & 99.94           & 99.00  & 99.99           & \textbf{100} \\
     & Tr OT  & \textbf{89.51}  & 48.85  & 88.81           & 0            \\
     & Te A* & 99.96           & 99.40  & \textbf{100}    & \textbf{100} \\
     & Te OT  & \textbf{89.36}  & 48.80  & 88.92           & 0            \\
    \midrule
    \multirow{4}{*}{WL Diff (\%)}
     & Tr A* & \textbf{-0.810} & -0.100 & -0.786          & 0   \\
     & Tr OT  & \textbf{-0.310} & +0.610 & -0.226          & {-} \\
     & Te A* & \textbf{-0.842} & -0.130 & -0.800          & 0   \\
     & Te OT  & \textbf{-0.352} & +0.590 & -0.244          & {-} \\
    \midrule
    \multirow{4}{*}{Via Diff (\%)}
     & Tr A* & -0.960          & +0.080 & \textbf{-1.042} & 0   \\
     & Tr OT  & -0.066          & +1.130 & \textbf{-0.118} & {-} \\
     & Te A* & \textbf{-0.952} & +0.040 & -0.928          & 0   \\
     & Te OT  & \textbf{-0.316} & +1.110 & -0.312          & {-} \\
    \midrule\midrule
    \multicolumn{6}{l}{\textbf{Average Runtime Comparison}} \\
    \midrule
    \multicolumn{2}{l}{Method} & \textbf{Constrained A*} & \multicolumn{2}{c}{\textbf{Proposed}} & \textbf{OR-Tools} \\
    \midrule
    \multicolumn{2}{l}{Time (s)} & $\approx$ 0.03 & \multicolumn{2}{c}{$\approx$ 3} & $\approx$ 130 \\
    \bottomrule
    \end{tabular}
    
    \raggedright
    \footnotesize
    \vspace{4pt}
    \centering
    Tr: Train data, Te: Test data, A*: Constrained A*, OT: OR-Tools,\\ FMR: FM routing order, FMH: FM heuristic, MR: Manhattan routing order, MH: Manhattan heuristic.
\end{table}

\clearpage

\subsection{Case Study}\label{subsec4.3}

To further verify the generalization capability of the proposed method, experiments are conducted on more challenging routing instances. The key specifications of this dataset are summarized in Table~\ref{tab6}. Specifically, in Case 1, we increased the problem’s complexity by nearly doubling the number of nets included while maintaining the same grid size and edge capacity as the previous dataset. In Case 2, we increased the grid size, the number of nets, and the edge capacity; in particular, we significantly increased the number of nets, the edge capacity, and the corresponding time limit of the solver to better reflect the difficulty of real-world problems. The maximum expansion size during inference using the FM cost map was kept at 50,000, and we also examined whether the proposed method could solve problems that could not be solved within the time limit. Finally, by using the pretrained network model as-is without retraining, we directly verified not only the time efficiency but also the method’s ability to solve complex problems along with generalization capability of the trained network.

\clearpage

\subsubsection{Case 1: Net expansion with pretrained grid size}

\renewcommand{\arraystretch}{1.3}
\begin{table}[t]
    \centering
    \caption{Problem settings for different cases.}
    \label{tab6}

    \begin{tabular}{lcc}
    \toprule
    \multirow{2}{*}{Metric} & \multicolumn{2}{c}{Method} \\
    \cmidrule(l){2-3}
     & Case 1 & Case 2 \\
    \midrule
    \multirow{1}{*}{Initial edge capacity (x, y, z)}
     & (1, 1, 1) & (5, 5, 5) \\
    \multirow{1}{*}{Data size (constrained A*/CP-SAT/Failed-Timeout)}
     & (50/50/50) & (10/10/10) \\
     \multirow{1}{*}{Max expansion size per net (generation)}
     & 200,000 & 50,000 \\
     \multirow{1}{*}{Max expansion size per net (inference)}
     & 50,000 & 50,000 \\
     \multirow{1}{*}{Grid size per data (x, y, z)}
     & (30, 30, 2) & (35, 35, 2) \\
     \multirow{1}{*}{Number of nets per data}
     & 30 - 35 & 130 \\
     \multirow{1}{*}{Time limit for CP-SAT}
     & 30 min & 12 h \\
    \bottomrule
    \end{tabular}
\end{table}

\renewcommand{\arraystretch}{1.3}
\begin{table}[t]
    \centering
    \caption{Performance and runtime comparison in Case 1.}
    \label{tab7}

    \begin{tabular}{lcccc}
    \toprule
    & & Success Rate (\%) & WL Diff (\%) & Via Diff (\%) \\
    \midrule
    & A* & 98.4   & -1.496  & -0.997 \\
    & OT  & 93.6 & -0.104  & +1.010  \\
    & FT  & 74.4 & {-}     & {-}    \\
    \midrule\midrule
    \multicolumn{5}{l}{\textbf{Average Runtime Comparison}} \\
    \midrule
    \multicolumn{2}{l}{Method} & \textbf{Constrained A*} & \textbf{Proposed} & \textbf{OR-Tools} \\
    \midrule
    \multicolumn{2}{l}{Time (s)} & $\approx$ 0.06 & $\approx$ 6 & $\approx$ 1060 \\
    \bottomrule
    \end{tabular}
    
    \raggedright
    \footnotesize
    \vspace{4pt}
    \centering
    FT: Failed-timeout.
\end{table}

\begin{figure}
\centering
 \includegraphics[width=0.8\textwidth]{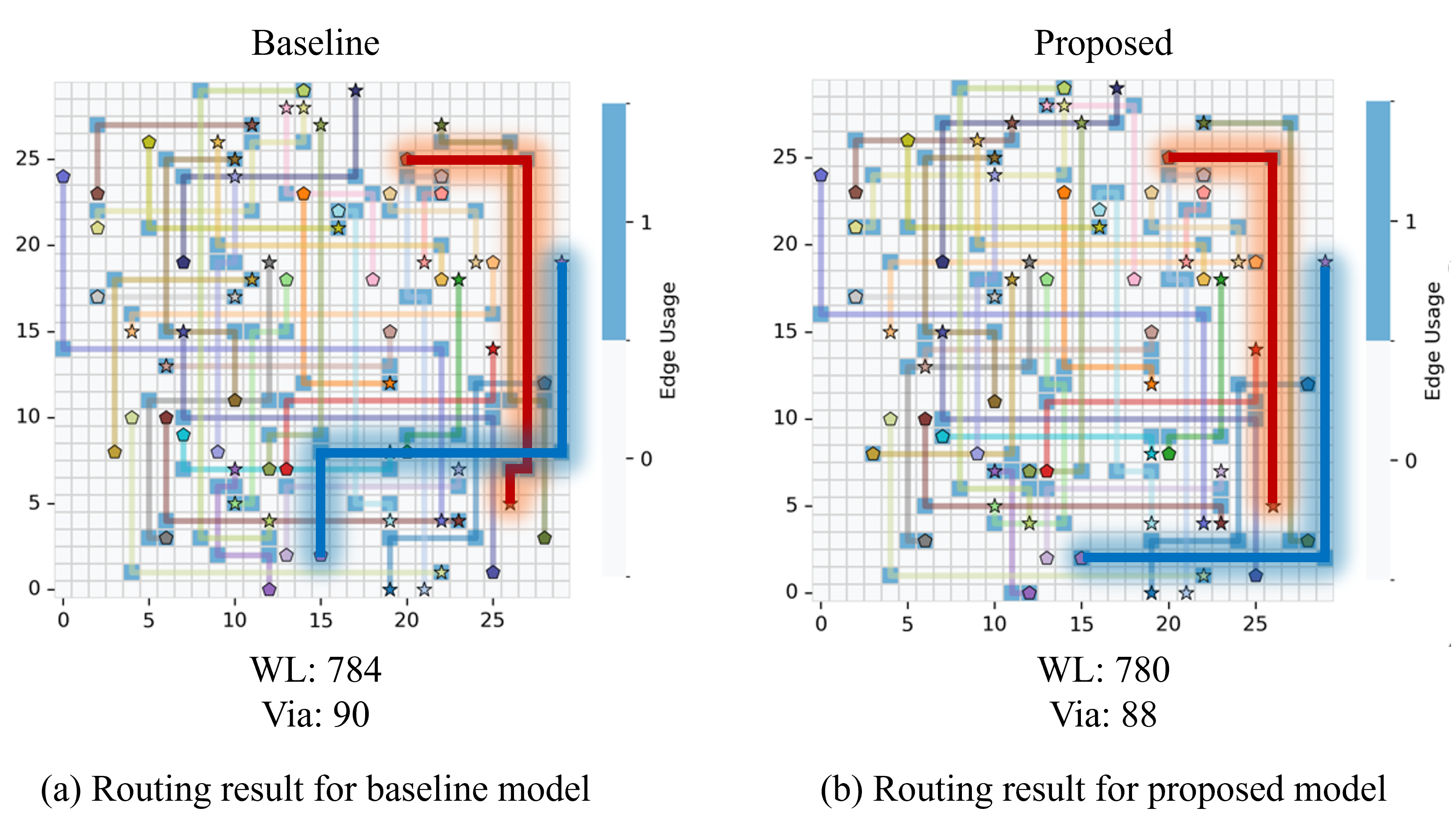}
\caption{Routing results comparison in Case 1.}
\label{fig:Success comparison2}
\vspace{-9pt}
\end{figure}

The result in Case 1 demonstrate that the proposed method achieves competitive performance, highlighting the effectiveness of the proposed algorithm for generating cost map in guiding the search process.
Detailed results are summarized in Table~\ref{tab7} and also displayed through qualitative comparisons in Fig.~\ref{fig:Success comparison2}. Overall, the proposed method maintains a high routing success rate even under increased problem complexity, demonstrating strong generalization capability.

For instances successfully solved by the solver, a slight increase in via count is observed compared to the baseline. However, this degradation is marginal when considering the significantly reduced inference time and the fact that the pretrained model was not exposed to such complex scenarios during training.
Additionally, the method attains a routing success rate of approximately 75\% on instances that remain unsolved by solver-based approaches within the given time limit. This result demonstrates the proposed framework's ability to handle challenging routing scenarios beyond the reach of traditional optimization methods.

\clearpage

\subsubsection{Case 2: Net expansion with untrained grid size}

As mentioned earlier, Case 2 uses data that more closely reflects real-world scenarios, and the runtime generally increases exponentially as the problem scale grows when using numerical solvers. Furthermore, it is worth noting that in this study, given that the S/T pins of all nets other than the routed net act as obstacles, the difficulty of these problems does not simply increase in proportion to the number of nets or the increase in edge capacity.

\begin{figure}
\centering
 \includegraphics[width=0.87\textwidth]{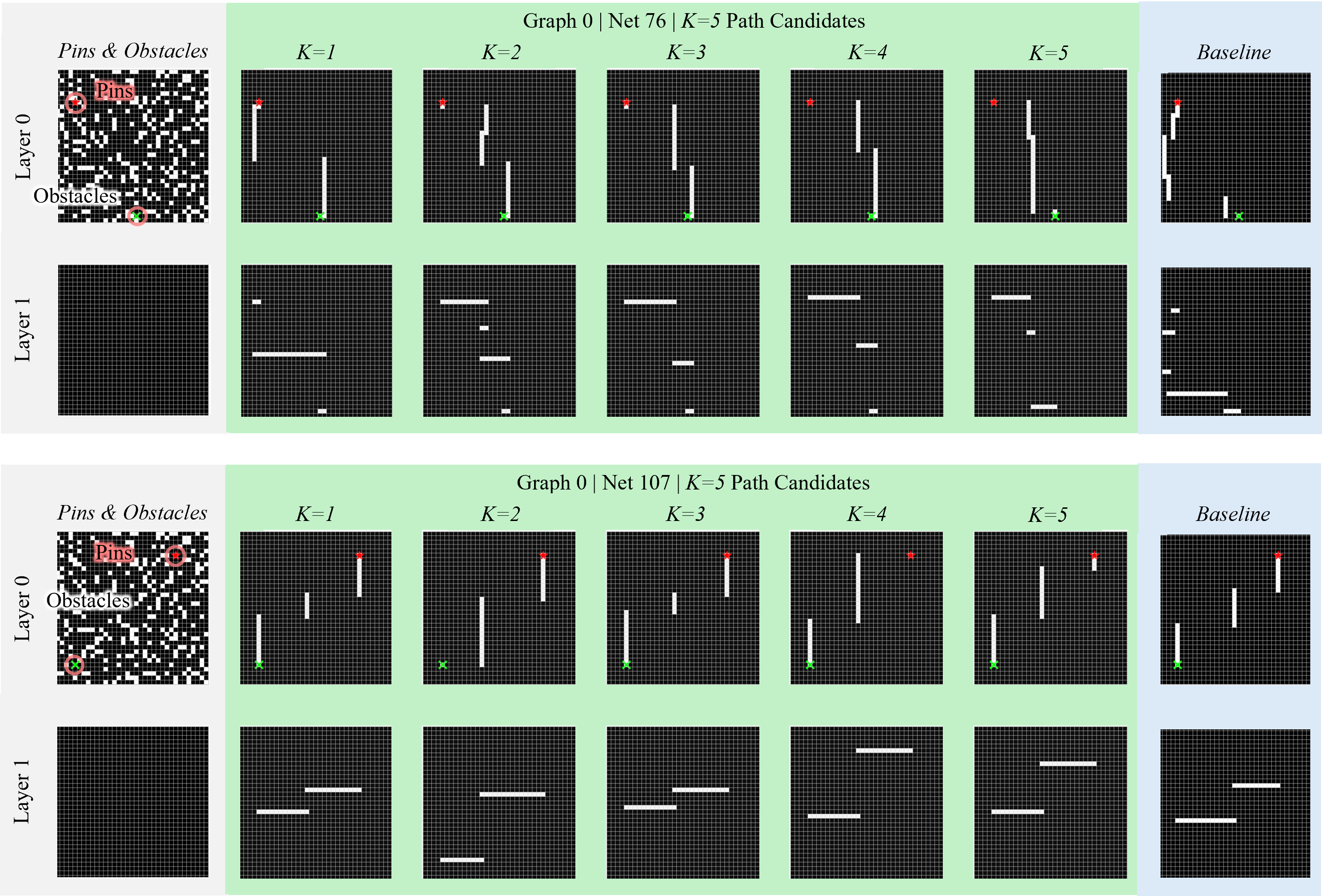}
\caption{Output results from 5 different noise sources in Case 2.}
\label{fig:noise_35}
\vspace{-9pt}
\end{figure}

\begin{figure}
\centering
 \includegraphics[width=0.8\textwidth]{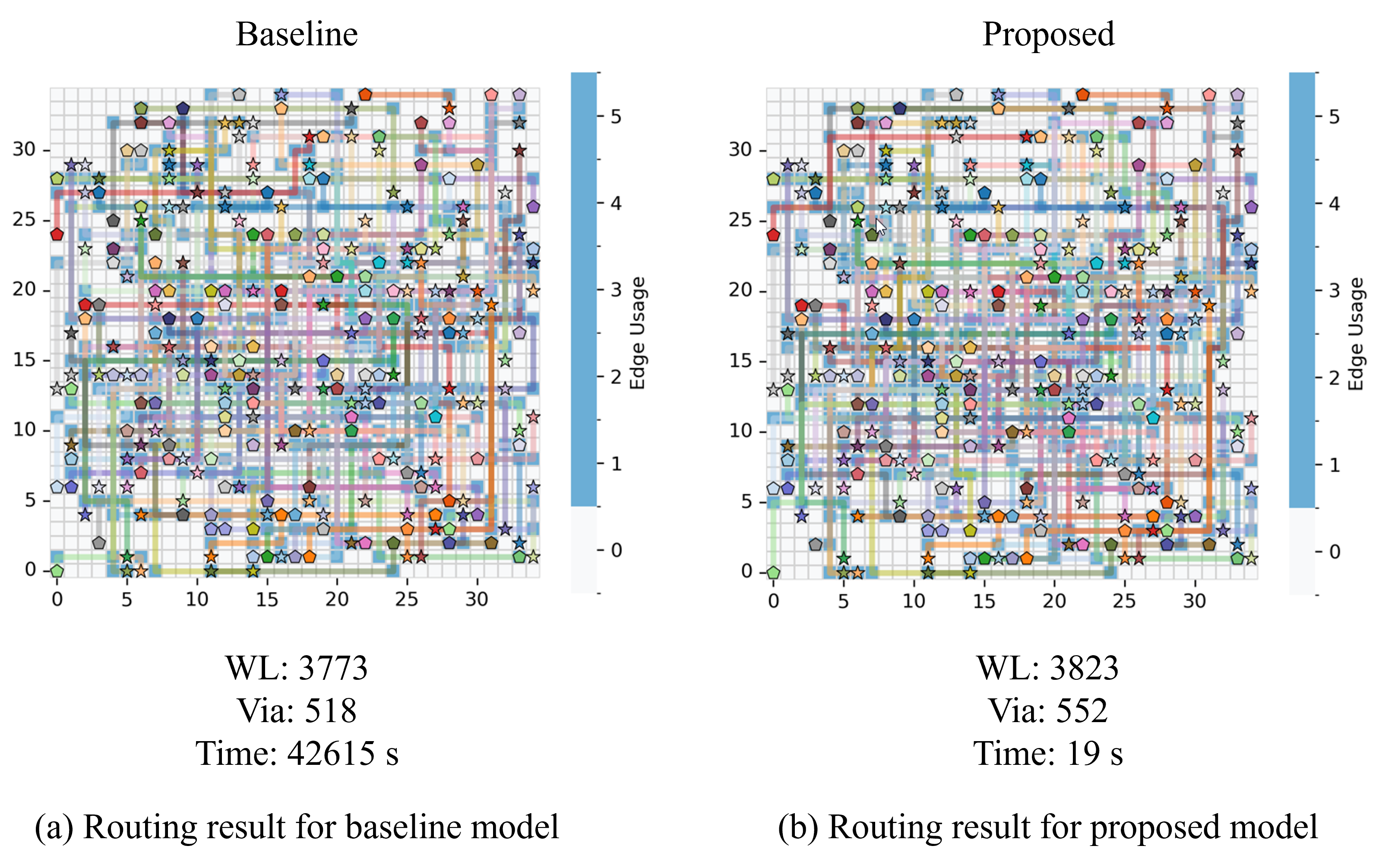}
\caption{Routing results comparison in Case 2.}
\label{fig:Success comparison3}
\vspace{-9pt}
\end{figure}

We began by examining the results of binary cost map generation with $K = 5$ different noise samples for the corresponding data, considering that the grid size is not in the trained range. The results are shown in Fig.~\ref{fig:noise_35}, which demonstrates that a binary cost map without mode collapse and satisfying connectivity requirements can be generated even for untrained grid size. Meanwhile, the specific validation results are shown in Table~\ref{tab8}, which confirms that the proposed methodology achieves a high success rate in this case as well. However, for problems solved using the solver, we observe that the values for WL and via count are higher than those of the baseline and have increased compared to Case 1. Nevertheless, as in the analysis for Case 1, this increase is reasonable when considering various factors, particularly the time required to solve the problem. comparison results between the baseline using the solver and the proposed method for a specific problem are illustrated in Fig~\ref{fig:Success comparison3}.

\renewcommand{\arraystretch}{1.3}
\begin{table}[h]
    \centering
    \caption{Performance and runtime comparison in Case 2.}
    \label{tab8}

    \begin{tabular}{lcccc}
    \toprule
    & & Success Rate (\%) & WL Diff (\%) & Via Diff (\%) \\
    \midrule
    & A* & 100   & -1.253  & -0.132 \\
    & OT  & 100 & +1.688  & +5.633  \\
    & FT  & 74 & {-}     & {-}    \\
    \midrule\midrule 
    \multicolumn{5}{l}{\textbf{Average Runtime Comparison}} \\ 
    \midrule
    \multicolumn{2}{l}{Method} & \textbf{Constrained A*} & \textbf{Proposed} & \textbf{OR-Tools} \\
    \midrule
    \multicolumn{2}{l}{Time (s)} & $\approx$ 0.03 & $\approx$ 18 & $\approx$ 17,780 \\
    \bottomrule
    \end{tabular}
\end{table}

\clearpage

\clearpage


\section{Conclusion}\label{sec5}

This study presents a cost-guided global routing framework that integrates FM-based cost map generation with constrained A*-based routing. The proposed approach leverages the generated cost maps to guide the search process, improving both routing feasibility and efficiency.

To the best of our knowledge, this is the first work to introduce binary FM for cost map generation in global routing and to systematically incorporate it into both routing order determination and heuristic design. Extensive experiments, including ablation studies and generalization tests, demonstrate that the proposed method achieves high success rates with competitive WL and via usage, while significantly reducing computation time compared to solver-based approaches. Furthermore, the method exhibits strong generalization performance, successfully solving a substantial portion of complex instances that conventional methods fail to solve within practical time limits.

Despite these advantages, several limitations remain. Routing failures still occur in highly congested scenarios, and the current framework does not explicitly model inter-net dependencies during cost map generation. In addition, the experiments are limited to 2-pin nets and a two-layer structure, and validation on industrial-scale datasets is not yet included.

Future work will focus on addressing these limitations by developing models that better capture inter-net relationships, extending the framework to multi-pin and multi-layer routing problems, and validating the performance on real-world datasets. Additionally, exploring fully end-to-end routing approaches that eliminate the reliance on heuristic search remains an important direction.


\section*{Acknowledgments} 
\paragraph{Funding}
This work was supported by grants from the Ministry of Science and ICT 
(GTL24031-000, N10250154, and No.~2022-0-00986), the Ministry of Trade, Industry and Energy 
(RS-2025-02317327 and RS-2025-25444634), and the research fund for newly appointed professors of Jeonbuk National University in 2026.

\section*{Declarations}
The authors declare that they have no known competing financial interests or personal relationships that could influence the work reported herein.

\section*{Replication of results}
The code and data are available from the corresponding author on reasonable request.


\printcredits

\bibliographystyle{cas-model2-names}

\bibliography{cas-refs}

\vskip3pt
\newpage




Namwoo Kang is an Associate Professor of CCS Graduate School of Mobility at KAIST. He is also currently CEO of Narnia Labs. Prior to joining KAIST, he served as a Research Fellow in the Department of Mechanical Engineering at the University of Michigan and worked as a Research Engineer at Hyundai Motor Company. He has earned his Ph.D. in Design Science at the University of Michigan. Previously, he obtained a M.S. degree in Technology and Management and B.S. in Mechanical and Aerospace Engineering from Seoul National University. He has been pursuing Generative AI-driven Product Design research by integrating physics and data for virtual product development. His research interests include generative design, data-driven design, machine learning, deep learning, design optimization, topology optimization, CAD/ CAM/CAE, and HCI.


\end{document}